%% file: imc2016.tex
\documentclass[10pt]{sig-alternate-05-2015}

\usepackage{ae}
\usepackage{cite}
\usepackage{amsmath}
\usepackage{amssymb}
\usepackage{graphicx}
\usepackage{booktabs}
\usepackage{tabularx}
\usepackage{supertabular,multicol}
\usepackage{balance}
\usepackage{microtype}
\usepackage{multirow}
\usepackage{color}
\usepackage{fancyvrb}
\usepackage[utf8]{inputenc}
\usepackage{dcolumn} %required from R generated tables
\usepackage[labelformat=simple]{subcaption}

\usepackage{xspace}

\usepackage{graphicx}
\usepackage{multirow}
\usepackage{hyperref}
\usepackage{tikz}

\usepackage{lipsum}

\title{Towards Better Internet Citizenship: \\ Reducing the Footprint of Internet-wide Scans by \\Topology Aware Prefix Selection}

\numberofauthors{4}
\author{
\alignauthor Johannes Klick\\
\affaddr{Freie Universit\"at Berlin}\\
\email{johannes.klick@fu-berlin.de}
\and
\alignauthor Stephan Lau \\
\affaddr{Freie Universit\"at Berlin}\\
\email{stephan.lau@fu-berlin.de}
\and
\alignauthor Matthias W\"ahlisch \\
\affaddr{Freie Universit\"at Berlin}\\
\email{m.waehlisch@fu-berlin.de}
\and
\alignauthor Volker Roth \\
\affaddr{Freie Universit\"at Berlin}\\
\email{volker.roth@fu-berlin.de}
}

\input{commands}

\begin{document}
\CopyrightYear{2016} 
\setcopyright{acmlicensed}
\conferenceinfo{IMC 2016,}{November 14 - 16, 2016, Santa Monica, CA, USA}
\isbn{978-1-4503-4526-2/16/11}\acmPrice{\$15.00}
\doi{http://dx.doi.org/10.1145/2987443.2987457}
\maketitle

\input{abstract}

\vspace{0.58cm}
\input{intro}
%\clearpage
\input{related}
%\input{tass}
\input{methodology}

\input{evaluation}

\input{discus_future}

\input{conclusion}
\input{acks}

\balance

\bibliography{literature}
\bibliographystyle{abbrv}

\end{document}

%% file: commands.tex
\usepackage{pifont}

\newcommand{\one}{({\em i})\xspace}
\newcommand{\two}{({\em ii})\xspace}
\newcommand{\three}{({\em iii})\xspace}

\makeatletter
\renewcommand{\paragraph}[1]{\vspace*{0.03in}\noindent{\bf #1.}\hspace{0.25ex \@plus1ex \@minus.2ex}}
\makeatother

%% file: abstract.tex
\begin{abstract}
Internet service discovery is an emerging topic to study the deployment of
protocols.
Towards this end, our community periodically scans the entire advertised
IPv4 address space.
In this paper, we question this principle.
Being good Internet citizens means that we should limit scan traffic to what
is necessary.
We conducted a study of scan data, which shows that several prefixes do not
accommodate any host of interest and the network topology is fairly stable.
We argue that this allows us to collect representative data by scanning
less. In our paper, we explore the idea to scan all prefixes once and then
identify prefixes of interest for future scanning.

Based on our analysis of the \emph{censys.io} data set (4.1 TB data encompassing 28
full IPv4 scans within 6 months) we found that we can reduce scan traffic
between 25-90\% and miss only 1-10\% of the hosts, depending on desired
trade-offs and protocols.

\end{abstract}

%% file: intro.tex
\section{Introduction}
\label{sec:intro}

Fast Internet-wide scanning is growing in popularity among researchers.  At
the time of writing, researchers regularly scan the Internet for vulnerable
SSL certificates~\cite{durumeric2013analysis,heninger2012mining}, SSH
public keys~\cite{gasser2014deeper}, and for the banners of plain text
protocols such as SMTP, HTTP, FTP, and Telnet~\cite{Durumeric:2015:SEB:2810103.2813703}.
%
% or services which can be used for amplification attacks e.g. open DNS resolver.
%
The majority of researchers scan at least 2.8 billion addresses advertised
in the IPv4 address space~\cite{%
        heidemann2008census, heninger2012mining,
        Durumeric:2015:SEB:2810103.2813703, durumeric2013zmap, durumeric2013analysis, eff_obs,
        nappa2014cyberprobe, leonard2010demystifying, gasser2014deeper,   shadowserver
        }.
Hitrates, the
fraction of probed addresses from which a response is received, are very
often under two percent~\cite{durumeric2013zmap}.  This means that most scan
traffic is overhead.  Most of these scans are done periodically for trend
analyses, which exacerbates the amount of unnecessary scan traffic.  For
example, the ongoing Internet-wide research project
\emph{censys.io}~\cite{Durumeric:2015:SEB:2810103.2813703,
durumeric2013zmap} probes the IANA allocated address space for 19 protocols
on a continuous basis. This results in 72.2 billion generated IP-packets
per week. which causes several hostile responses ranging from threatening legal actions 
to conducted denial-of-service attacks~\cite{durumeric2013zmap}.
Whereas scanning the IPv4 address space is feasible this is not any more the case
for IPv6.  When IPv6 becomes more popular, we need scanning strategies that
limit scans to parts of the address space that are in use.

Many measurement scenarios require only partial scans instead of
exploring the full IP address space. However, we currently lack a
systematic understanding of the deployment of Internet services with
respect to IP address ranges.

\iffalse
While scanning the IPv4 address space is feasible this is not any more the case
for IPv6.  When IPv6 becomes more popular, we need more efficient scanning strategies
which focus on addresses in use.
\fi

%In this paper, we present the \emph{Topology Aware Scanning Strategy (TASS)}, a new IP prefix based and topology aware scanning strategy for periodic scanning.  \emph{TASS} enables researchers to collect responses from 90-99\% of the available hosts for six months by scanning only 10-75\% of the announced IPv4 address space in each scan cycle (protocol dependent).

In this paper, we want to start the discussion how we can reduce scan traffic systematically.
We present the \emph{Topology Aware Scanning Strategy (TASS)}, a new IP prefix based and topology aware scanning strategy for periodic scanning.  \emph{TASS} enables researchers to collect responses from 90-99\% of the available hosts for six months by scanning only 10-75\% of the announced IPv4 address space in each scan cycle (protocol dependent).
\emph{TASS} is seeded with the results of a full advertised IPv4 address
scan for a given protocol and time period.  The prefixes for all responses
will be selected for periodic scans of the given protocol.

Periodic scanning of only selected prefixes reduces scan traffic
significantly while hitting most of the hosts of interest.  For instance,
our analysis reveals that responsive prefixes obtained from a full FTP scan cover 98\% of all FTP
hosts 6 months later, at the cost of scanning only 57.4\% of the advertised
addresses.  The scanning overhead can be optimized further by omitting
prefixes with a low \emph{density.}  Here, \emph{density} denotes the
fraction of hosts per address space size.  For example, if we limit prefix
selection to a 95\% coverage of the responsive addresses then we can still
find 92.3\% of the FTP hosts after six months while scanning only 20.6\% of
the announced address space. Moving forward we plan to investigate whether
the distributions of found and missed hosts are the same.

For our evaluation of \emph{TASS} we use 4.1\,TB of data derived from 28
full IPv4 scans obtained from
\emph{censys.io}~\cite{Durumeric:2015:SEB:2810103.2813703}.  For
common protocols we show that, following an initial scan of the full IPv4 address space, the hitrate for
responsive prefixes decreases by about 0.3 percent per month compared to
what a full scan would find.

Consequently, periodical \emph{TASS} scans
are 1.25 to 10 times more efficient for a period of at least 6 months if
researchers accept a single-digit percentage reduction in host coverage.

% Local Variables:
% mode: latex
% coding: utf-8
% TeX-master: "imc2016"
% End:

%% file: related.tex
\section{State of the Art}
\label{sec:related-work}

% \begin{figure}%[ht]
%   \centering
%   \includegraphics[width=\columnwidth]{images/related.pdf}
%   \caption{Current scanning strategies and their scoping of the address space.}
%   \label{fig:related}
% \end{figure}

\begin{figure}
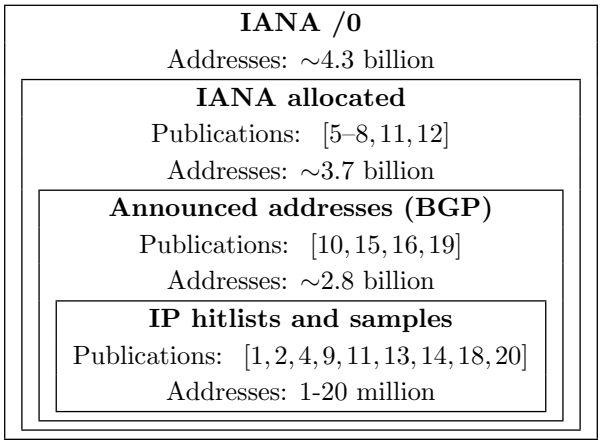

  \centering
  \renewcommand{\arraystretch}{1.2}
  \begin{tabular}{|c|}
    \hline
    \textbf{IANA /0}\\
    Addresses: $\sim$4.3 billion\\
    \begin{tabular}{|c|}
      \hline
      \textbf{IANA allocated}\\
      Publications:
     ~\cite{%
        heidemann2008census, heninger2012mining,
        Durumeric:2015:SEB:2810103.2813703, durumeric2013zmap, durumeric2013analysis, eff_obs   }\\
      Addresses: $\sim$3.7 billion\\
      \begin{tabular}{|c|}
        \hline
        \textbf{Announced addresses (BGP)}\\
        Publications:
       ~\cite{%
          nappa2014cyberprobe, leonard2010demystifying, gasser2014deeper,
          shadowserver     }\\
        Addresses: $\sim$2.8 billion\\
        \begin{tabular}{|c|}
          \hline
          \textbf{IP hitlists and samples}\\
          Publications:
         ~\cite{%
            alt2014uncovering, rossow2014amplification, fan2010selecting,
            cai2011understanding, heidemann2008census, holz2011ssl, lee2007cryptographic, cicalese2015characterizing, yilek2009private     }\\
          Addresses: 1-20 million\\
          \hline\noalign{\smallskip}
        \end{tabular}\\
        \hline\noalign{\smallskip}
      \end{tabular}\\
      \hline\noalign{\smallskip}
    \end{tabular}\\
    \hline\noalign{\smallskip}
  \end{tabular}
  \caption{Current scanning strategies and their scoping of the IPv4 address
    space.}
  \label{fig:related}
\end{figure}

% Many large-scale Internet-wide scanning studies have been conducted in the
% past~\cite{ heidemann2008census, heninger2012mining,
% Durumeric:2015:SEB:2810103.2813703, durumeric2013zmap,
% leonard2010demystifying, gasser2014deeper, alt2014uncovering,
% shadowserver,rossow2014amplification}.  This section will analyze these
% projects regarding efforts that had been taken for decreasing the IPv4
% address space being scanned.

\emph{TASS} represents a trade-off between scanning overhead and results
accuracy.  In what follows, we review the kinds of trade-offs other
researchers have made previously.  We identified three kinds of approaches
in the literature: \one full scans of IANA allocated addresses, \two scans
of routable addresses and \three scans of address space samples.

\paragraph{IANA allocated address space}
The most basic approach is to scan all IP addresses covered by the /0
prefix.  Scans of this type seek to explore the reachability of all
potential hosts.  However, some (unicast) addresses do not offer public
services per definition, for example, private networks addresses and the
loopback addresses.  Excluding these unallocated or reserved addresses is
the first obvious step towards a reduction of scanning noise.  This has been
a common practice from the beginning
\cite{  heidemann2008census, heninger2012mining, durumeric2013analysis, eff_obs  } and is still being practiced
\cite{Durumeric:2015:SEB:2810103.2813703, durumeric2013zmap}.

% \todo{Does any of these scans require to scan addresses that are not
%   advertised in BGP (e.g., they consider incomplete BGP visibility)? We need
%   to make clear (a) what they want to achieve, be precise, and (b) how they
%   approach their goal. Then, we can hopefully argue that a full scan is not
%   always required.}

\paragraph{Announced IP addresses (BGP)}
The second type of trade-off involves only addresses that are covered in
global BGP tables~\cite{nappa2014cyberprobe, leonard2010demystifying, gasser2014deeper,
 shadowserver}.
 
%
% It is worth noting that BGP tables are not complete in the sense that they
% include all possible paths towards a destination but they should ensure
% reachability to this destination.  In the authors used a complete
% BGP-table for their scanning activities and reduced the IPv4 address space
% to 2.6-2.7 billion IP addresses.
%

% \todo{Why does this approach not belong to Hitlists? It looks like a hybrid
%   version.}

\paragraph{IP hitlists and samples}
Several researchers sampled parts of the IPv4 address space in order to
extrapolate from their data.  For example, Alt \emph{et
  al.}~\cite{alt2014uncovering} scanned for honeypots by probing at least
one host in all /24 blocks of the Internet.
Rossow~\cite{rossow2014amplification} used a random sample of~1 million IP
addresses in his research on traffic amplification threats.  Heidemann
\emph{et al.}~\cite{heidemann2008census} probed 1\% of the address space
repeatedly, which consisted of 24,000 /24 blocks.  These blocks were
compiled based on three different selection strategies: \one 50\% were
selected randomly, \two 25\% were selected if a host in this block was
responsive before, and \three 25\% were selected by other policies.  This
approach does not discriminate between prefixes of different sizes and
therefore it does not utilize potentially important topology information.

Sampling leads to a reduction of scan traffic, but is less suited for research
that requires precise statistics.
Whereas samples tend to be probabilistic, hitlists are compiled based on
predetermined characteristics.  Fan and Heidemann~\cite{fan2010selecting}
generated IP address hitlists by scanning the IPv4 address space repeatedly
and by filtering out addresses that were consistently responsive.  Their
approach was applicable to only a third of the Internet, though, and
exhibited 40-50\% fluctuation after three months, probably caused by dynamic
IP addresses.  By comparison, \emph{TASS} compiles prefix hitlists and
exhibits only 1-10\% fluctuation after six months.  Dynamic IP addresses
fluctuate within a particular prefix, which may explain why \emph{TASS} is
significantly more stable.

Cai and Heidemann~\cite{cai2011understanding} investigated the
responsiveness of /24 blocks (note the difference from /24 network \emph{prefixes}).
They probed~1\% of the Internet address space
by selecting /24 blocks that were responsive to ICMP probes, as shown by a
prior census of all allocated addresses.  They clustered blocks with
adjacent addresses and similar network behavior, and found that a fifth of
the /24 blocks had a utilization less than 10\%.

\iffalse
Several researchers use the Alexa listings~\cite{alexa} for web protocols scans or
other host lists derived from network traffic observations or scanning
activities~\cite{holz2011ssl,lee2007cryptographic,yilek2009private}.
\fi

Plonka \emph{et al.}~\cite{ plonka2015temporal} used passive
IPv6 measurements of WWW clients for identification of stable and dense IPv6 prefixes.
We use active scan data and focus on hosts,
but we are aware that a combination of both approaches
might lead to a more comprehensive solution.

\paragraph{Summary}
The objectives of most of the measurement studies do not require a priori
scanning of unreachable address space.
The state of the art in Internet scanning appears to base trade-offs
primarily on IP blocks and individual IP addresses.  We are not aware of
attempts other than ours to leverage network prefix responsiveness for
scan traffic reduction.

% Local Variables:
% mode: latex
% coding: utf-8
% TeX-master: "imc2016"
% End:

%% file: methodology.tex
\clearpage

\section{Topology-Aware Scanning}

In this section, we give a high-level overview over TASS, followed by an
empirical motivation why TASS is a promising trade-off between scanning
overhead and accuracy.  An evaluation of TASS performance over time is given
in Section~\ref{sec:eval}.  We used the FTP, HTTP, HTTPS, and CPE WAN Management Protocol (CWMP).
For brevity, we provide graphs primarily for FTP and HTTPS.

% This section describes the the core idea of this paper, the host to prefix
% distribution of the Internet for some protocols and the procedure of further
% data enrichement with prefixes provided by Routeviews Prefix-to-AS mappings
%~\cite{caidapfx2as}.

\input{tass}

\subsection{Prefix derivation}
\label{sec:prefixes}

\emph{TASS} requires that addresses are mapped to prefixes.  The
\emph{censys.io} dataset already contains prefix information that Durumeric
\emph{et al.}~\cite{Durumeric:2015:SEB:2810103.2813703} apparently obtained
from their outgoing AS. However, closer inspection reveals that the
included information is often coarse-grained or even missing.  For this
reason, we chose to use the Routeviews Prefix-to-AS mappings (pfx2as)
provided by CAIDA~\cite{caidapfx2as} instead.  Said mappings reflect a
topological view of the Internet, are fine-grained, and are used routinely
for research. 

It is worth noting that prefixes in BGP may be loosely aggregated. 
In particular, more specific prefixes ($m$-prefixes, e.g. 100.0.0.0/12) may be announced in parallel to less specific prefixes ($l$-prefixes, e.g.  100.0.0.0/8).
The CAIDA data includes a large fraction of more specific prefixes in addition to less specific prefixes.
For example, the dataset of 2015/09/07 contains 595,644 prefixes of which 54\% are $m$-prefixes.
The $m$-prefixes account for 34.4\% of the advertised IP space.

%
% (less and more specific prefixes; 322,282 in 595,644 prefixes or
% 967,568,713 of 2,810,606,010 IPs, as of 2015/09/07).
%

\begin{figure}
  \centering
  \begin{subfigure}[b]{0.48\columnwidth}
  \centering
  {\footnotesize
   \newcommand\bstart{0} % x start value of blocks
    \begin{tikzpicture}[scale=0.8]
      %\draw[step=1cm,gray,very thin] (0,0) grid (4,5);
      
      \draw (\bstart+0,0) rectangle (\bstart+4,4);
      \draw[dashed] (\bstart+0,3) -- (\bstart+1,3);
      \draw[dashed] (\bstart+1,3) -- (\bstart+1,4);

      %\fill[green] (\bstart+0,2) -- (\bstart+1,2) -- (\bstart+1,4) -- (\bstart+0,4) -- cycle;

      \node[align=center] at (\bstart+2,4.5) {$l$-prefix /8};
      \draw[>=latex,->,] (\bstart+1, 1.2) -- (\bstart+0.5, 3.3);
      \node[align=right] at (\bstart+1.2, 1) {$m$-prefix};
      
      \node[align=center] at (\bstart+2,2) {/8};
      \node[align=center] at (\bstart+0.5,3.5) {/12};
      
    \end{tikzpicture}}
    \caption{Announced prefixes.}
  \end{subfigure}
  ~
  \begin{subfigure}[b]{0.48\columnwidth}
  \centering
  {\footnotesize
   \newcommand\bstart{0} % x start value of blocks
    \begin{tikzpicture}[scale=0.8]
      %\draw[step=1cm,gray,very thin] (0,0) grid (4,5);
      
      \draw (\bstart+0,0) rectangle (\bstart+4,4);
      \draw (\bstart+2,2) -- (\bstart+2,4);
      \draw (\bstart+0,2) -- (\bstart+4,2);
      \draw (\bstart+1,2) -- (\bstart+1,4);
      \draw (\bstart+0,3) -- (\bstart+1,3);

      %\fill[green] (\bstart+0,2) -- (\bstart+1,2) -- (\bstart+1,4) -- (\bstart+0,4) -- cycle;

      \node[align=center] at (\bstart+2,4.5) {$l$-prefix /8};
      %\draw[->] (\bstart-0.5, 3) -- (\bstart+0.2, 3);
      %\node[align=right] at (\bstart-1.2, 3) {$m$-prefix};
      
      %\node[align=center] at (2,4.5) {/8};
      \node[align=center] at (\bstart+2,1) {/9};
      \node[align=center] at (\bstart+3,3) {/10};
      \node[align=center] at (\bstart+1.5,3) {/11};
      \node[align=center] at (\bstart+0.5,3.5) {/12};
      \node[align=center] at (\bstart+0.5,2.5) {/12};
      
    \end{tikzpicture}}\\
    \caption{Resulting $m$-prefixes.}
  \end{subfigure}
  \caption{The less specific $l$-prefix /8 contains the more
    specific $m$-prefix /12. The $l$-prefix is then decomposed
    into the more specific one and the remaining smallest prefixes.}
  \label{fig:iptree}
\end{figure}
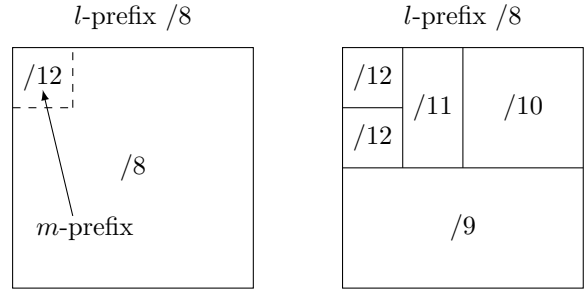

To reflect potential network characteristics, we deaggregate the $l$-prefix of each $m$-prefix into the minimal set of prefixes that contains the $m$-prefix.
This approach allows us to take all routing information into account 
while maintaining a proper partition of the address space for scanning purposes.  
See Figure~\ref{fig:iptree} for an illustration of this process. 
In the following two sections we show that this approach potentially reduces the number of
scanned addresses.

\subsection{Host stability versus prefix length}
\label{prefix_analysis}

\begin{figure*}
  \centering
  \begin{subfigure}[b]{0.98\columnwidth}
    \includegraphics[width=\textwidth]{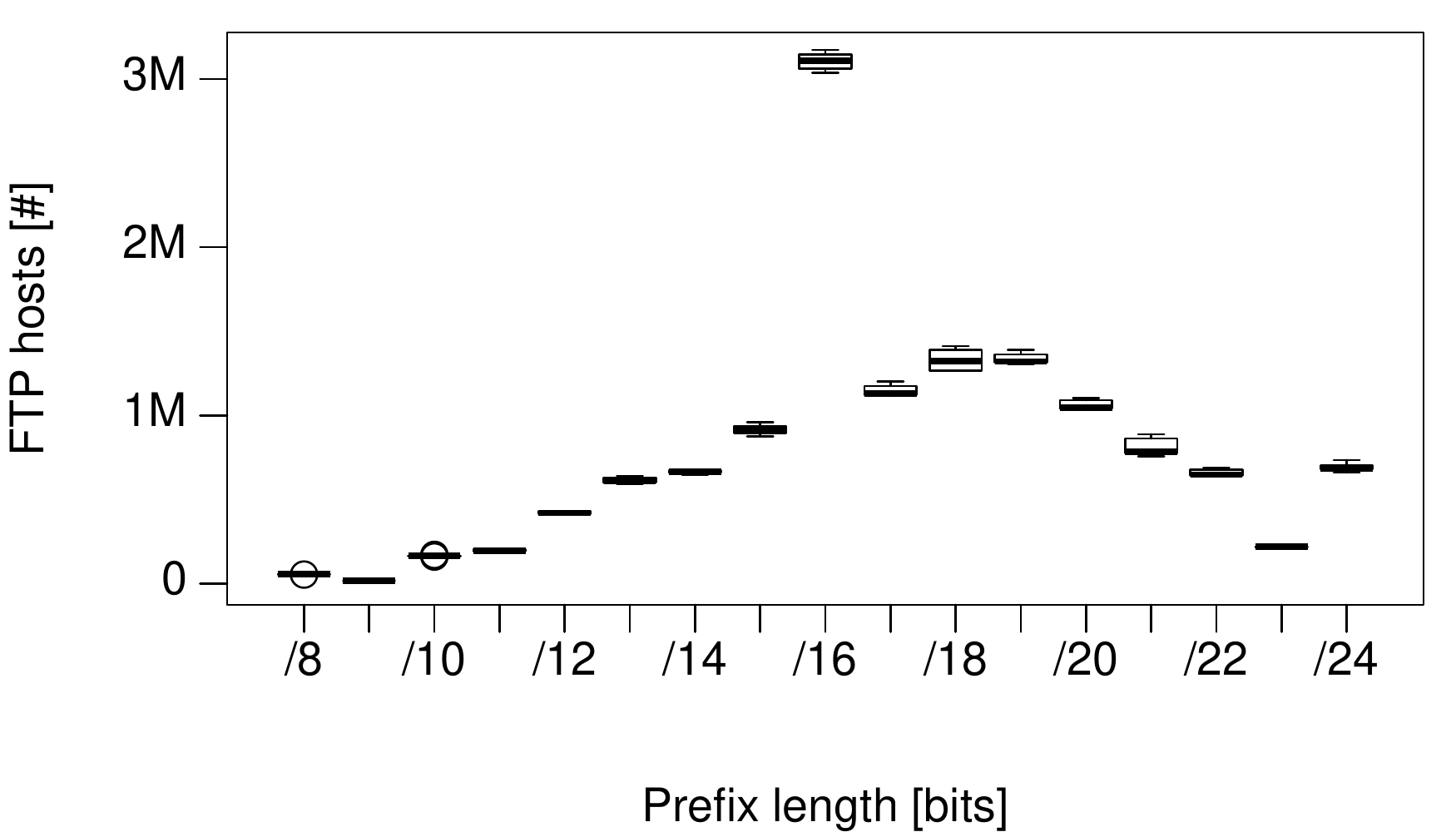}
    \caption{FTP for less specific prefixes.\\[1em]}
    \label{fig:less_specific_dist_ftp}
  \end{subfigure}
  ~
  \begin{subfigure}[b]{0.98\columnwidth}
    \includegraphics[width=\textwidth]{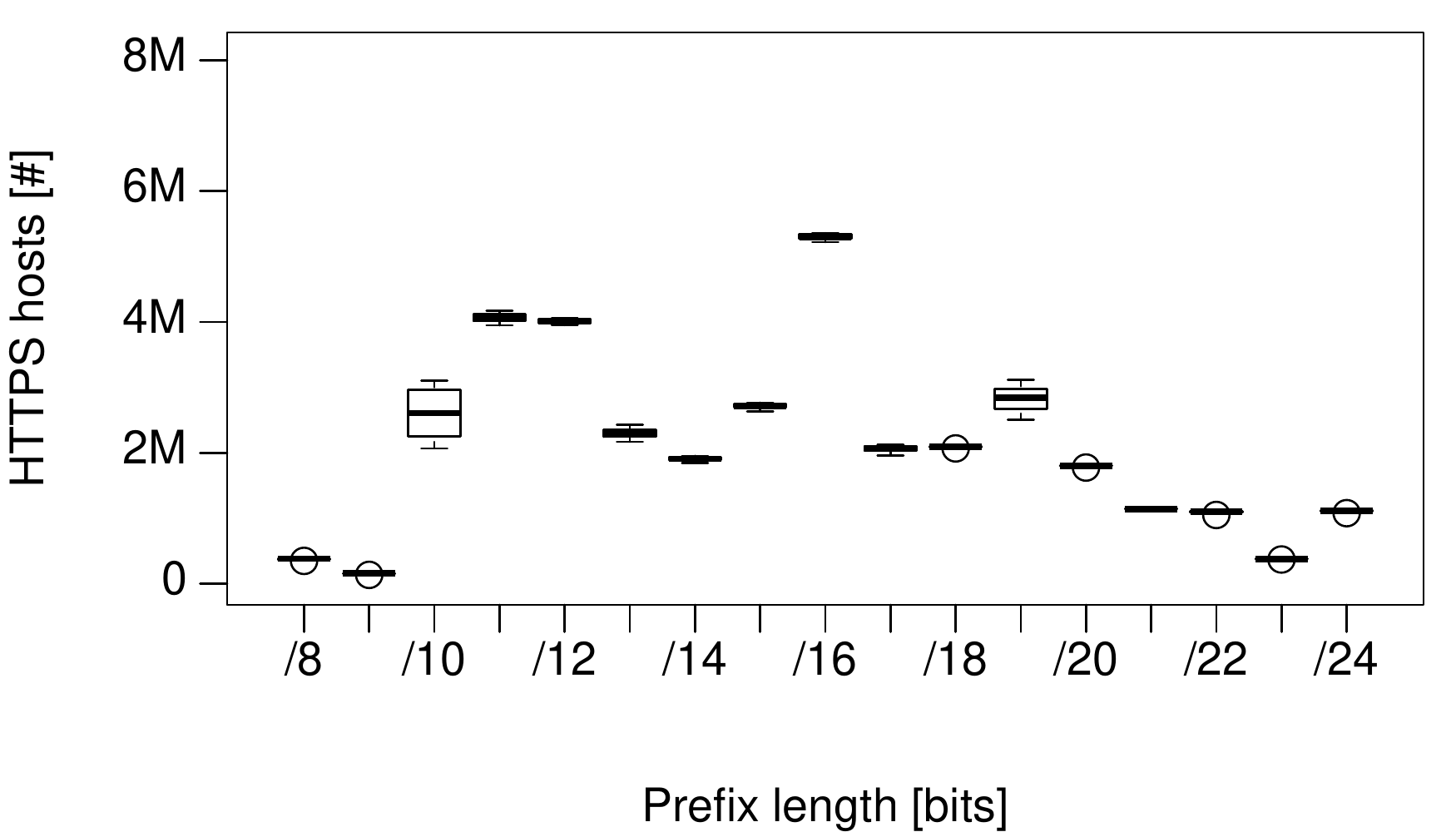}
    \caption{HTTPS for less specific prefixes.\\[1em]}
    \label{fig:less_specific_dist_https}
  \end{subfigure}
  
  \begin{subfigure}[b]{0.98\columnwidth}
    \includegraphics[width=\textwidth]{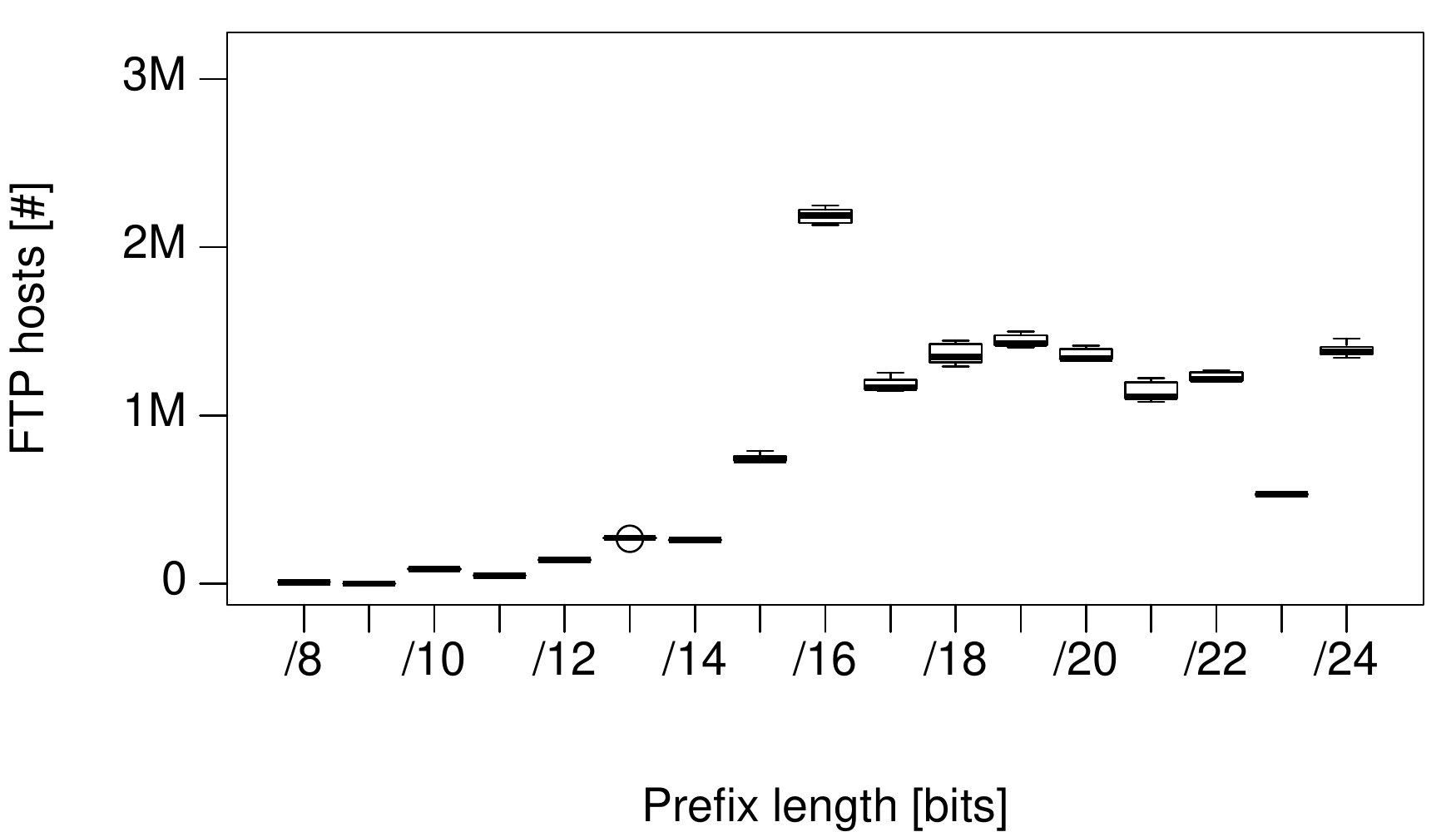}
    \caption{FTP for more specific prefixes.}
    \label{fig:more_specific_dist_ftp}
  \end{subfigure}
  ~
  \begin{subfigure}[b]{0.98\columnwidth}
    \includegraphics[width=\textwidth]{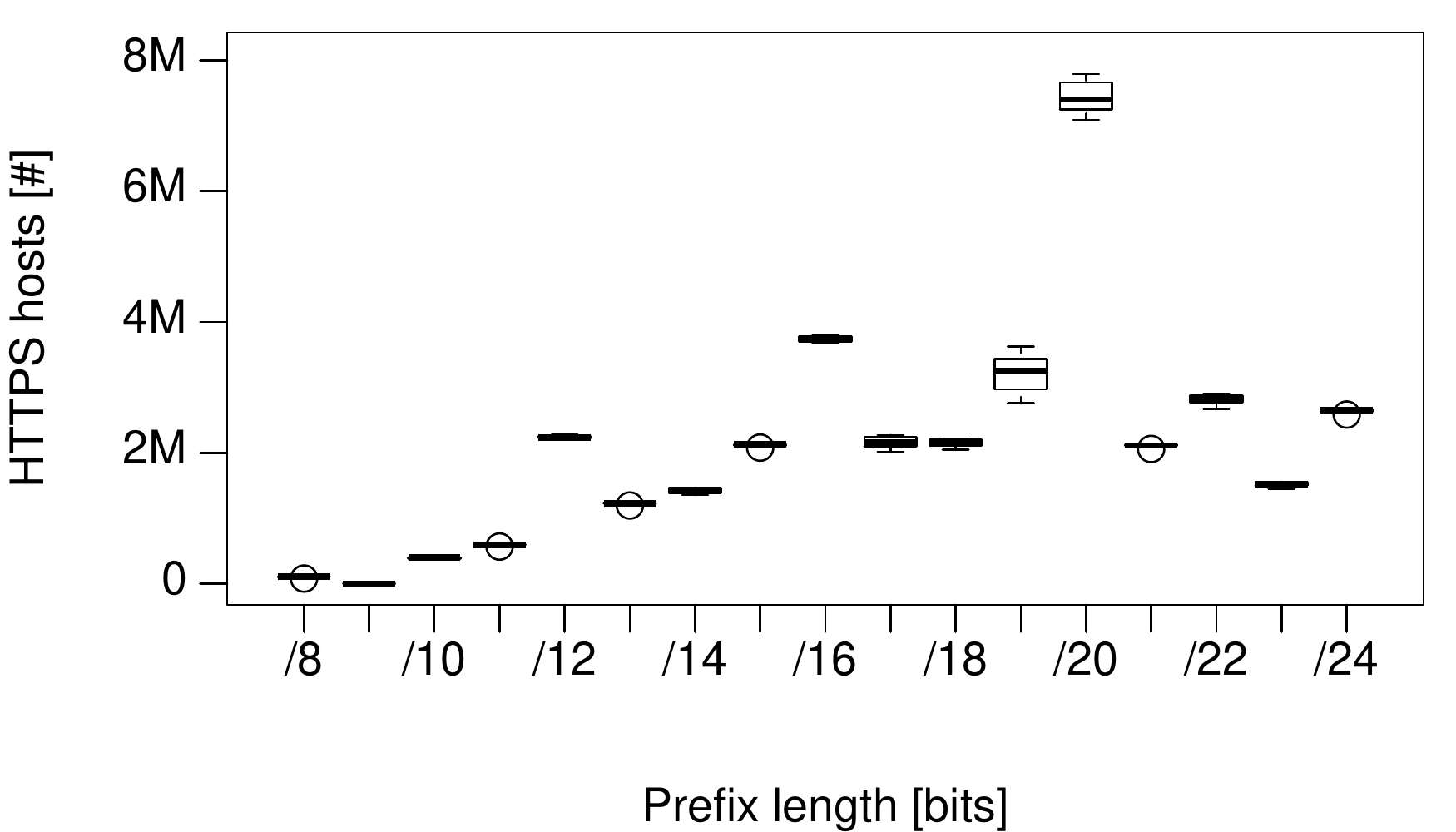}
    \caption{HTTPS for more specific prefixes.}
    \label{fig:more_specific_dist_https}
  \end{subfigure}
  \caption{Shows the host distribution over prefix lengths based on seven different measurements
    from 09/2015 to 03/2016. Prefixes longer than /24 are negligible and
    have been omitted.}
  \label{fig:prefix_distributions}
\end{figure*}

We expect that TASS performs well if hosts do not fluctuate significantly in
between prefixes.  In a first step, we analyzed the distribution of host
numbers across prefixes of different lengths over a period of six months
with 7~measurements.  If the distribution variance was high then this would
already indicate that TASS may miss hosts.
Figures~\ref{fig:less_specific_dist_ftp}
and~\ref{fig:less_specific_dist_https} show the results for the case of
$l$-prefixes, and Figures~\ref{fig:more_specific_dist_ftp}
and~\ref{fig:more_specific_dist_https} show the results for the case of
$m$-prefixes.  The host numbers appear to be stable and therefore the results
do not contradict our expectation.  Of course, this result is necessary but
not sufficient by itself.  We still need to investigate the fluctuation in
between prefixes of the same length.  This is future work that we intend to
do with a larger dataset and for a full paper.  The graphs also indicate a
right-shift towards longer prefixes without a pronounced loss of stability.
This lends support to our hypothesis that $m$-prefixes are a better choice
than $l$-prefixes because their density is potentially higher.  For example,
if all hosts in an $l$-prefix cluster in an $m$-prefix then the $l$-prefix minus
the $m$-prefix need not be scanned.

\begin{figure*}
  \centering
  \begin{subfigure}[b]{\columnwidth}
    \includegraphics[width=\textwidth]{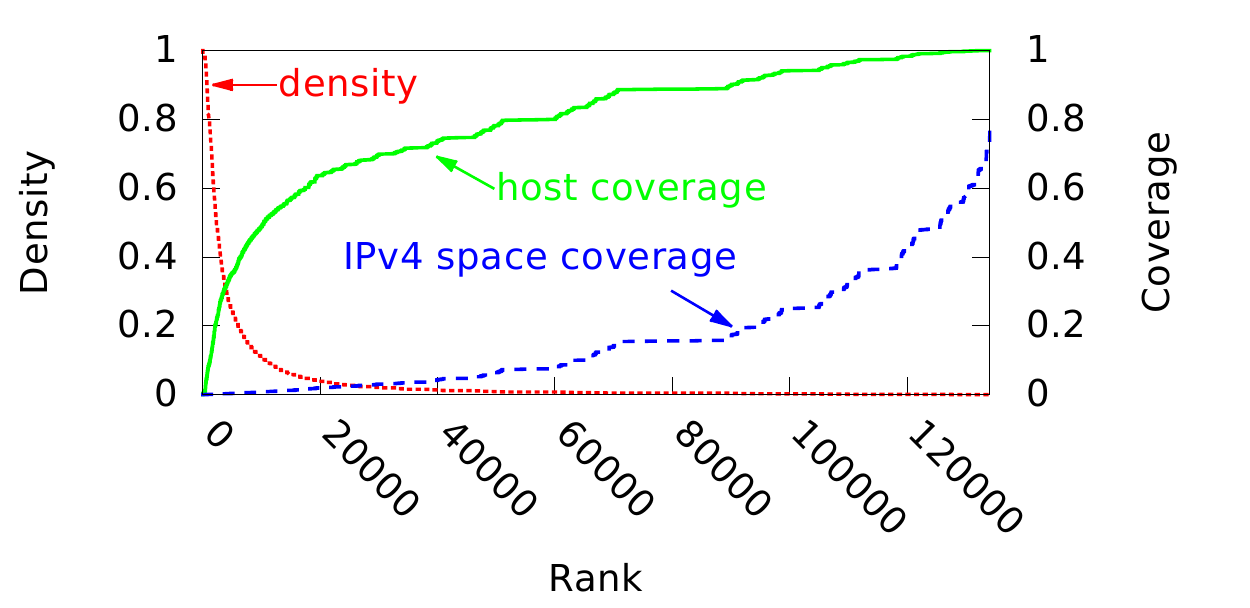}
    \caption{FTP: Less specific prefixes.\\[1em]}
    \label{fig:l_prefix_density_ftp}
  \end{subfigure}
  ~
  \begin{subfigure}[b]{\columnwidth}
    \includegraphics[width=\textwidth]{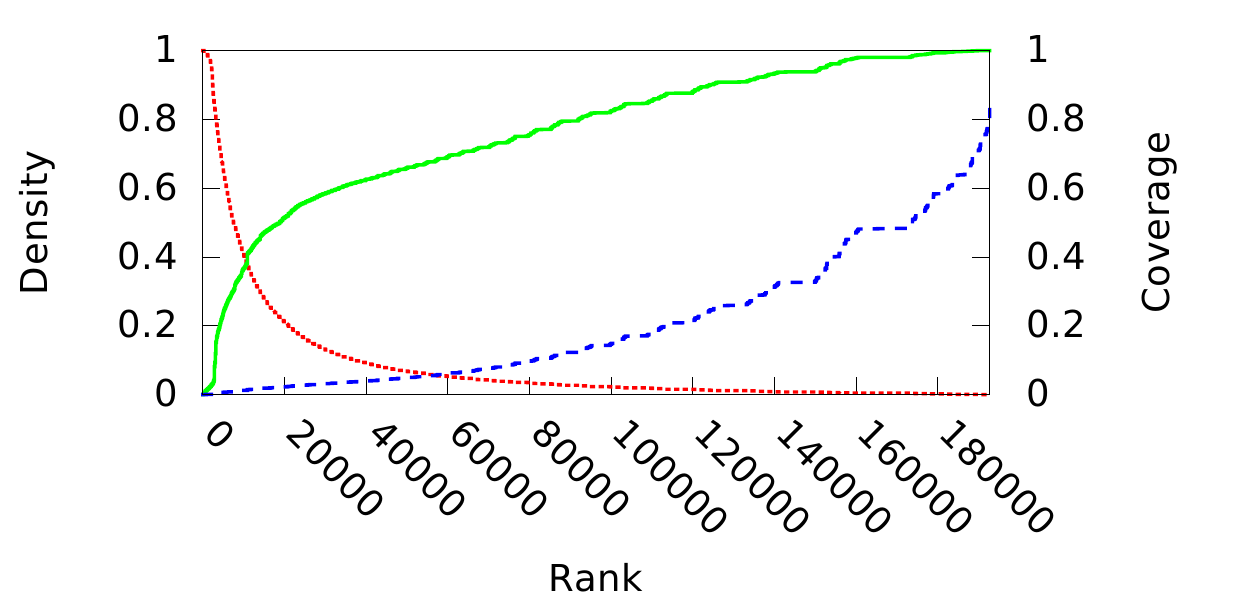}
    \caption{HTTP: Less specific prefixes.\\[1em]}
    \label{fig:l_prefix_density_http}
  \end{subfigure}

  \begin{subfigure}[b]{\columnwidth}
    \includegraphics[width=\textwidth]{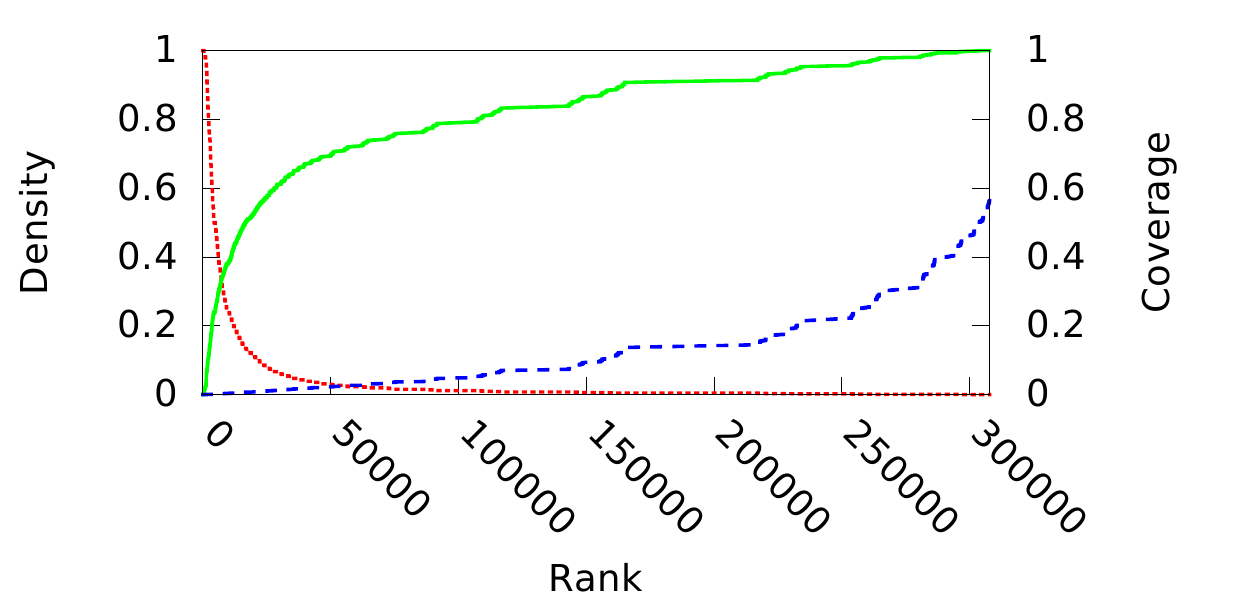}
    \caption{FTP: More specific prefixes.}
    \label{fig:m_prefix_density_ftp}
  \end{subfigure}
  ~  
  \begin{subfigure}[b]{\columnwidth}
    \includegraphics[width=\textwidth]{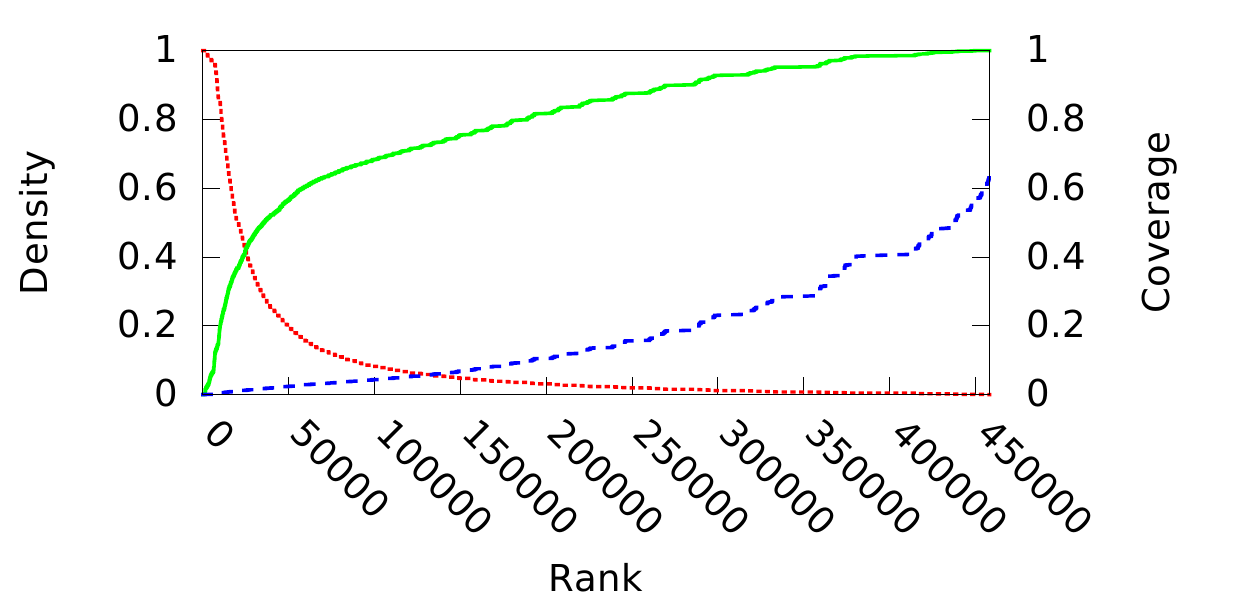}
    \caption{HTTP: More specific prefixes.}
    \label{fig:m_prefix_density_http}
  \end{subfigure}
  %\caption{The dotted curve shows the prefixes ranked by
    %their density. The solid curve shows cumulative relative host coverage and
    %dashed shows the cumulative relative address space coverage.}%
  \caption{Shows responsive prefixes ranked by their density (dotted), the cumulative
  relative host coverage (solid), and the cumulative relative address space
  coverage (dashed) with density $\rho>0$.}%
  \label{fig:prefix_densities}%
\end{figure*}

\iffalse

Die Abb. \ref{prefix_size} zeigt die Verteilung von FTP Host abgetragen über
die verschienden zugehörigen Prefix-Größen.  Es wurden pro Monat von 09/15
bis 03/16 monatlich ein Datensatz ausgewählt.  Der Boxplot zeigt, dass die
Verteilung für ein halbes Jahr relativ stabil ist.  Desweiteren scheinen
Protokolle in einem ähnlichen Anwednungsgebiet wie z.B. HTTP, HTTPS und FTP
eine ähnliche Verteilung aufzuzeigen.  Wohingegen Protokolle wie Telnet und
CWMP andere Verteilungen ausweisen.  Mithilfe der empirischen
Kullback-Leibler-Distanz wird ein Maß der verschiedenheit zwischen den
einzelnen Verteilungen berechnet.  Sobald die Daten fertig sind, erwarte ich
dass für FTP, HTTP und HTTPS sehr geringe unterschiede aus der
Kullback-Leibler-Distanz hervorgehen, sodass man die Gruppieren könnte.

Next we can show that for each protcol exists a uniqe distribution in
relation to the IP's prefix size.  Fig \ref{prefix_size} shows the
distrinution of FTP hosts over their prefix size.

show protcol distribution per fix prefixes
prefix distribution is protocol specific

\fi

\subsection{Prefix density}
\label{sec:prefix_density}

TASS yields a favorable trade-off if small reductions in coverage lead to
large reductions in scan overhead.  Towards an evaluation of the potential
trade-off, we analyzed the density of prefixes in relation to the advertised
address space.  Recall that the density of a prefix is the number of
responsive hosts in the prefix divided by the number of addresses in the
prefix.  Figures~\ref{fig:l_prefix_density_ftp}
and~\ref{fig:l_prefix_density_http} show the results for the case of
$l$-prefixes, and Figures~\ref{fig:m_prefix_density_ftp}
and~\ref{fig:m_prefix_density_http} show the results for the case of
$m$-prefixes.  The graphs are sorted in the order of decreasing prefix density
$\rho$, the red curve.  Prefixes with zero density are not included.  The
green curve is the \emph{cumulative relative host coverage} $\phi$.  The
blue curve is the \emph{cumulative relative address space coverage.}  The
graphs show a sharp decrease of prefix density combined with a sharp
increase in host coverage and a modest increase of address space coverage
over the range of prefixes.  This clearly indicates that prefix selection
based on prefix density is well suited to maximize the efficiency of scans.
Based on the data we analyzed we can report the following statistics for the
case of $l$-prefixes:
\begin{itemize}
\item 100\% ($\phi=1$) of all FTP hosts are found in $\sim$134\,K prefixes
  representing 76.2\% of the routed address space.
\item 95\% ($\phi=0.95$) of all FTP hosts are found in $\sim$105\,K prefixes
  representing 27.3\% of the routed address space.
\item 23.8\% of the addresses were unresponsive.
\item The first 20\,K prefixes with a density of $\rho>0.04$ contain 64\%
  ($\phi=0.64$) of all FTP servers but represent only 2\% of the advertised
  address space.
\end{itemize}
\begin{table}
  \begin{tabular}{ lll l l l l }
\toprule
                                                       &                                        & $\phi$ & FTP   & HTTP  & HTTPS & CWMP \\
  \cmidrule{3-7}\cmidrule{3-7}
  \multirow{8}{*}{\rotatebox{90}{Address Space Coverage}}  & \multirow{4}{*}{\rotatebox{90}{less}} & 1      & 0.762 & 0.828 & 0.832 & 0.477\\
                                                       &                                        & 0.99   & 0.470 & 0.548 & 0.542 & 0.142\\
                                                       &                                        & 0.95   & 0.273 & 0.362 & 0.343 & 0.099\\
                                                       &                                        & 0.7    & 0.031 & 0.064 & 0.065 & 0.043\\
                                                       &                                        & 0.5    & 0.008 & 0.021 & 0.024 & 0.024\\
  \cmidrule{2-7}
                                                       & \multirow{4}{*}{\rotatebox{90}{more}} & 1      & 0.574 & 0.648 & 0.645 & 0.332\\
                                                       &                                        & 0.99   & 0.371 & 0.440 & 0.427 & 0.113\\
                                                       &                                        & 0.95   & 0.206 & 0.279 & 0.262 & 0.085\\
                                                       &                                        & 0.7    & 0.023 & 0.048 & 0.052 & 0.037\\
                                                       &                                        & 0.5    & 0.006 & 0.017 & 0.020 & 0.021\\
\bottomrule
  \end{tabular}
\caption{IPv4 address space coverage of the protocols using less and more specific prefixes.}
\label{tab:phi}
\end{table}

For $m$-prefixes, an address space coverage of 57.4\% suffices to achieve full
host coverage ($\phi=1$), which is a reduction of 18.8 percentage points
compared to $l$-prefixes.  At the same time, prefix selection based on density
is roughly twice as efficient as a full scan, for the FTP protocol.  If one
tolerates a 5\% loss of FTP hosts then scanning 20.6\% of the address space
suffices to find 95\% of the FTP hosts that a full scan would find.
For detailed information, see Table~\ref{tab:phi}.

\iffalse
Es wird pro Präfix die Hostdichte pro spezifischen Protkoll berechnet.
Die Dichte brechnet sich aus Anzahl der Hosts in einem Prefix dividiert durch die Anzahl der IP-Adressen die durch den Prefix repräsentiert werden.
Abbildung \ref{prefix_density} zeigt die Prefixe sortiert nach ihrer Dichte für FTP.
Die grüne Linie stellt eine CDF Funktion über gefunden Hosts dar.
Die Blaue Linie stellt eine CFF Funktion über die Menge von IP-Adressen dar, die durch alle Prefixe repräsentiert werden.
Folgende Aussagen lass sich aus  \ref{prefix_density} ablesen:
\begin{itemize}
    \item 100 Prozent aller FTP Host sind in 57,4 Prozent des gerouteten IPV4-Adresspaces.
    \item 95 Prozent aller FTP hosts sind in 20,7 Prozent  des gerouteten IPV4-Adresspaces zu finden.
\end{itemize}

Datendatei: 20150907\_ipBlock\_FTP.sql.csv

Man könnte folglich 95 Prozent aller FTP Hosts bei einem TASS SCAN erhalten
und dabei mit ungefähr fünffacher Effizienz scannen.

\fi

% we will analyze the for HTTP and FTP. (DNS, HTTPS, Telnet?)  onversely,
% some researcher scan a random subset and extrapolate their results in the
% following~\cite{rossow2014amplification}.  or trend analyses they need to
% perform periodically scans.

% Local Variables:
% mode: latex
% coding: utf-8
% TeX-master: "imc2016"
% End:

%% file: tass.tex
\hfill
\subsection{TASS in a Nutshell}
\label{sec:tass}

\emph{TASS} amortizes the overhead of an initial scan of the full routable
address space over repeated scans that cover only a subset of all
prefixes. The core idea of \emph{TASS} is to identify prefixes which are of
primary interest when scanning the Internet repeatedly.  The goal of
\emph{TASS} is to be \emph{efficient.}  Efficiency is measured as the number
of successful protocol handshakes per number of connection attempts.
\emph{TASS} is parameterized by an adjustable target ratio $\phi$
that specifies the proportion of hosts that \emph{TASS} shall cover in
repeated scans.  For this reason, we refer to $\phi$ as the host
\emph{coverage.} \emph{TASS} works as follows:
\begin{enumerate}
\item At time $t_0$, perform a full scan and output all responsive
  addresses.  Let $N$ be their number.  Count the number of responsive
  addresses $c_i$ in each responsive prefix $i$.  The sum of all $c_i$ is
  $N$.
\item Calculate the density $\rho_i=c_i/2^{32-\text{prefix length}}$ of all
  responsive prefixes and their relative host coverage $\phi_i=c_i/N$ of
  responsive addresses.
\item Sort the prefixes in the descending order of density.  Relabel
  prefixes so that $i<j \Leftrightarrow \rho_i > \rho_j$.
\item Find the smallest $k$ so that $\sum_{i=1}^k \phi_i > \phi$.
\item Scan prefixes $1, \dots, k$ repeatedly until time $t_0+\Delta_t$, then
  start over at step~1.
\end{enumerate}

Within each time interval $[t_0, t_0+\Delta_t]$ there will be a gradual loss
of results accuracy as hosts leave or enter prefixes other than prefixes $1,
\dots, k$.  On each full scan, full accuracy is recovered.  We motivate in
the remainder of this section why we expect this strategy to yield high
accuracy with significantly reduced scan overhead.  We evaluate the strategy
in Section~\ref{sec:eval} and quantify the expected loss of accuracy over
time, which yields an adjustable time period $\Delta_t$.

% Local Variables:
% mode: latex
% coding: utf-8
% TeX-master: "imc2016"
% End:

%% file: evaluation.tex
\section{Accuracy over time}
\label{sec:eval}
%
%\begin{figure}[ht]
%  \centering
%  \includegraphics[width=0.53\textwidth]{images/ftp_most/SUM_DEVS_20150907_.pdf}
%  \caption{FTP MOST: Result Stability: We show the hitrate of TASS over the time for 7 month in comparison to a full IPv4 scan. (FTP)}
%  \label{Stable_dev}
%\end{figure}
%
%\begin{figure}[ht]
%  \centering
%  \includegraphics[width=0.53\textwidth]{images/ftp_less/SUM_DEVS_20150907_.pdf}
%  \caption{FTP LESS Result Stability: We show the hitrate of TASS over the time for 7 month in comparison to a full IPv4 scan. (FTP)}
%  \label{Stable_dev1}
%\end{figure}
%
%
%\begin{figure}[ht]
%  \centering
%  \includegraphics[width=0.53\textwidth]{images/http_most/SUM_DEVS_20150915_.pdf}
%  \caption{HTTP MOST Result Stability: We show the hitrate of TASS over the time for 7 month in comparison to a full IPv4 scan. (FTP)}
%  \label{Stable_dev2}
%\end{figure}
%
%
%\begin{figure}[ht]
%  \centering
%  \includegraphics[width=0.53\textwidth]{images/http_less/SUM_DEVS_20150915_.pdf}
%  \caption{HTTP LESS Result Stability: We show the hitrate of TASS over the time for 7 month in comparison to a full IPv4 scan. (FTP)}
%  \label{Stable_dev3}
%\end{figure}
%

The findings we summarized in previous sections suggest that TASS can be an
efficient scanning strategy.  However, the benefits manifest only if the
distribution of hosts across prefixes remains reasonably stable over time.
As a first step to quantify the accuracy of TASS over time we simulated TASS
and an address-based hitlist approach using monthly snapshots of full IPv4
scans from \emph{censys.io}~\cite{Durumeric:2015:SEB:2810103.2813703} for
the time period from 09/2015 to 03/2016 (7~snapshots).  Then we determined
the fraction of hosts that TASS and the hitlist approach would have
uncovered in each scan cycle compared to a periodic full scan.  We used the
aforementioned datasets as our ground truth, again.  We focused our analysis
on four protocols, which were FTP, HTTP, HTTPS and TR-069 also known as the
CPE WAN Management Protocol (CWMP), a 4.1\,TB dataset in total.  CWMP is
used for remote management of residential gateways.
We chose CWMP for contrast because its purpose differs markedly from the
other two protocols.

\subsection{Hitlist accuracy over time}

\begin{figure}
  \centering
  \includegraphics[width=\columnwidth]{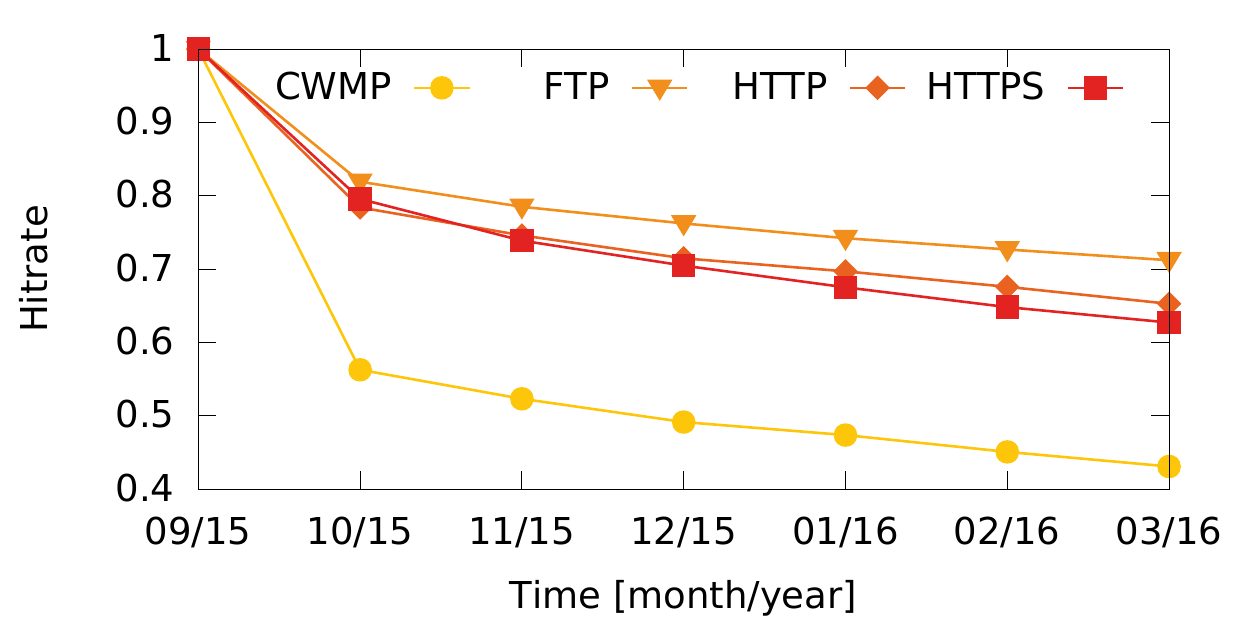}
  \caption{Hitrate using IP hitlists.}
  \label{fig:naive}
\end{figure}

The hitlist approach we simulated takes all addresses that are responsive in
an initial full scan and subsequently scans only those addresses.  This
strategy exhibits maximal efficiency and accuracy for stable (unchanging)
host distributions. Figure~\ref{fig:naive} show the results of our
simulation.  They indicate that the accuracy of the hitlist approach quickly
drops to 80\% within one month and continues to decrease over time for FTP,
HTTP and HTTPS.  The drop is much more pronounced for the CWMP protocol.  A
likely explanation is that residential gateways are connected to the
Internet via dynamic IP addresses more often.  Over the course of six
months, the accuracy drops to 71\% for HTTP and to 43\% for CWMP.  From
these results we conclude that the hitlist approach is not recommendable for
periodic scanning over time periods of several months.

\begin{figure*}[Htp!]
  \centering
  \begin{subfigure}[b]{\columnwidth}
    \includegraphics[width=\columnwidth]{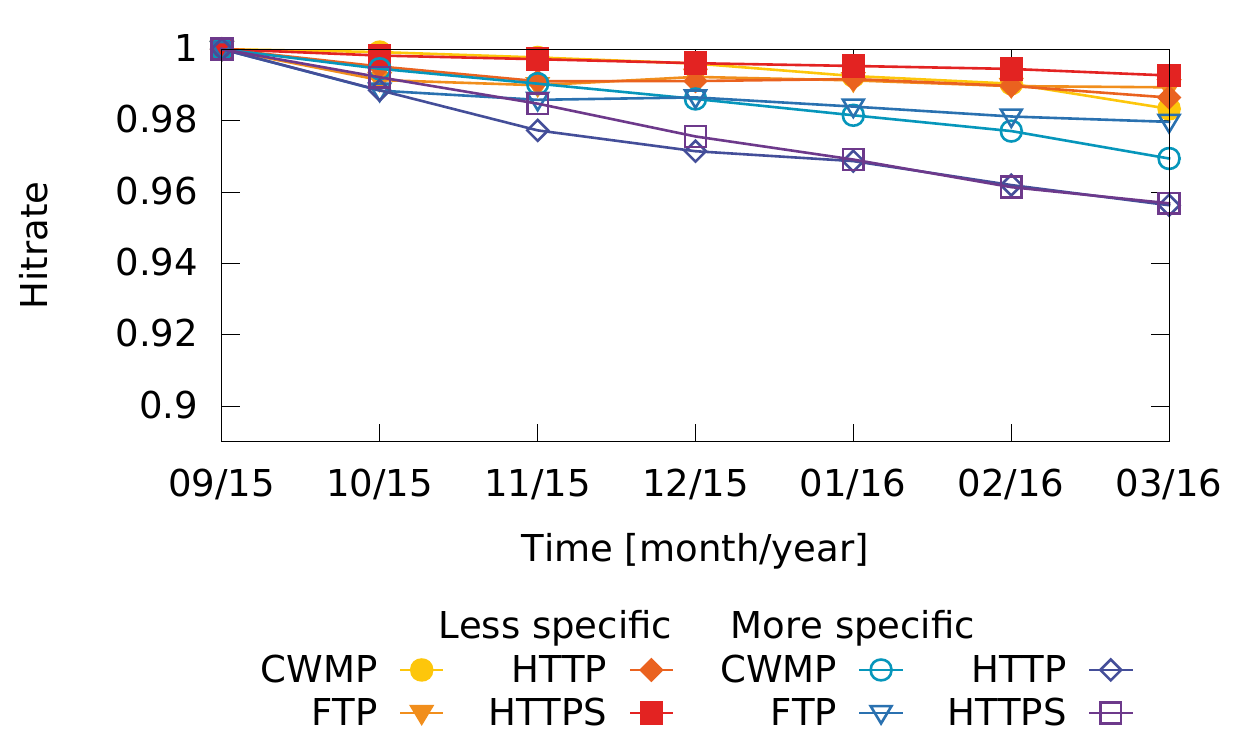}
    \caption{Hitrate with $\phi=1$.}
    \label{fig:tass_precision_100}
  \end{subfigure}
  \begin{subfigure}[b]{\columnwidth}
    \includegraphics[width=\columnwidth]{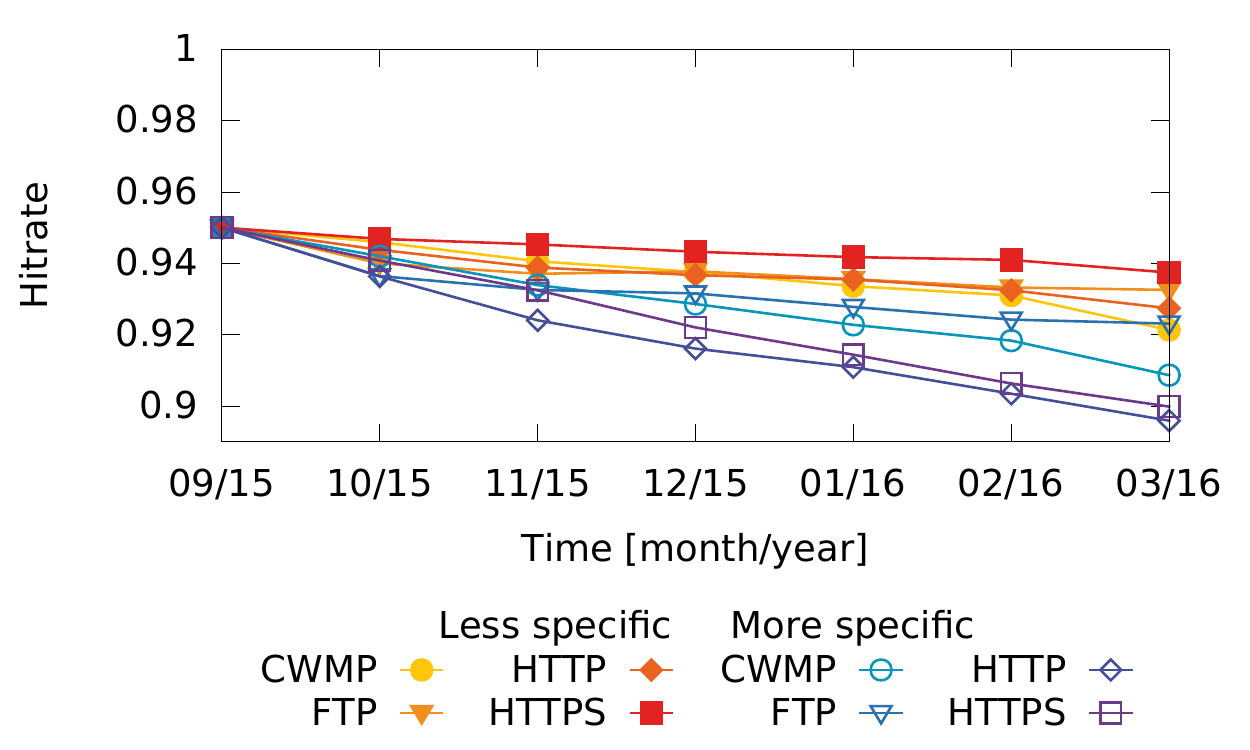}
    \caption{Hitrate with $\phi=0.95$.}
    \label{fig:tass_precision_95}
  \end{subfigure}
  \caption{Hitrate of TASS compared to a full scan.}
\end{figure*}

\subsection{TASS accuracy over time}

We simulated TASS with $l$-prefixes and $m$-prefixes as described prior over the
same six months time period. Figure \ref{fig:tass_precision_100} shows the
results for a coverage setting of $\phi=1$, that is, full host
coverage.  Recall that this selects all prefixes with a non-zero density,
that is, $\rho>0$.  We found that accuracy decreases at a rate of 0.3\% per
month for $l$-prefixes. For $m$-prefixes, accuracy decreases at a rate of up to
0.7\% per month or about 4.2\% over the course of six months.  The greater
efficiency of $m$-prefixes is thus paid for by an accuracy reduction twice as
much as for $l$-prefixes.  We repeated our analysis for a host coverage
setting of $95\%$, that is, $\phi=0.95$.  This reduced the accuracy further
to 90-94\%, depending on the protocol.  Figure~\ref{fig:tass_precision_95}
summarizes the outcomes.
We started a similar investigation of SSH and selected SCADA protocols but
to our surprise we found that accuracy and densities increased over time.
Further scrutiny of the ground truth datasets revealed that the snapshots
for these protocols likely included data from prior scans.  We have notified
the main contributor of
\emph{censys.io}~\cite{Durumeric:2015:SEB:2810103.2813703} who acknowledged
the problem.

\iffalse
Eine wichtige Frage ist, wie lange die initial ausgewählten Prefix stabil bleiben bzw. Hosts beinhalten.
Abbildung \ref{stable_dev} zeigt den zeitlichen Verlauf der Hitrate über 6 Monate.
Als Hitrate verstehen wir den Anteil der mit TASS gefundenen Geräte gegenüber aller durch censys.io gefundenen Geräte.
Wie man der Abbildung leicht entnehmen kann, wärem die duch TASS gesammelten Ergebnisse pro Monat ca. 0.3 Prozent schlechter als ein voller IPv4 Scan.
Wenn man dabei beachtet, dass TASS jedoch 80 Prozent weniger SCAN-Traffic erzeugt, ist dies ein akzeptables Ergebnis für eine neuartige Scan-Strategie.
For a good Internet citizenship :-)
\fi

\iffalse
Abb. \ref{stable_prefix} zeigt die Anzahl der durch TASS identifzierten Prefixe
 im Vergleich zu der Anzahl aller existierender Präfixe die 
bei einem vollen IPv4-Scan identifiziert worden wären. Man kann erkennen, dass nach 6 Monaten nur 90 Prozent aller Prefixe
die FTP Hosts beinhalten von TASS gescannt worden wären. 

TODO: Hier wäre sicherlich ehere zusätzlich noch interessant den Vergleich über die durch Präfix representierte IP-Adressmenge anzustellen. 

Welche weiteren Analysen wären aus deiner Sicht noch sinnvoll?

good for often repeated scans for ssh or https key material  like done in X Y Z
comparison table how many ip could have been saved from beeing scanned  
x times more efficient scan approach

\fi

% Local Variables:
% mode: latex
% coding: utf-8
% TeX-master: "imc2016"
% End:

%% file: discus_future.tex
\section{Discussion and Future Work}
\label{sec:disc_future}

Our results indicate that \emph{less} specific prefixes yield greater
scanning accuracy over time than \emph{more} specific prefixes.  A
likely cause is that $l$-prefixes reduce the overall number of prefixes, which
renders it less likely that a host fluctuates in between prefixes.  On the
other hand, $l$-prefixes have a higher scanning overhead compared to
$m$-prefixes.  For a full host coverage setting ($\phi=1$), the overhead
differed by about 15-20 percentage points according to our analysis in
Section~ \ref{sec:prefix_density}.  Consequently, we must consider this
trade-off when deciding between $l$-prefixes and $m$-prefixes.

Likewise, the host coverage setting $\phi$ has a significant influence on
the scanning overhead.  Even a small reduction of host coverage, say from
$\phi=1$ to $\phi=0.99$, results in a reduction of scan overhead by 20-30\%.
As part of our future work we intend to investigate more closely how the 1\%
of missed hosts are distributed in comparison to the other hosts.

Furthermore, in the context of the analysis of security incidents (e.g.,
\emph{Heartbleed}) it is important to analyse whether vulnerable servers
are distributed equally across both selected prefixes and omitted prefixes, for
$\phi<1$.  If the distribution was fairly equal then regular estimates of
vulnerable populations could be obtained with good efficiency and accuracy,
for example, with $\phi=0.5$ and a small address space coverage
of 0.6-0.8\% per scan cycle.

Finally, we suspect that more fine-grained prefixes may help to reduce
the scanning overhead even further.  Towards this end, it may be worthwhile
to apply the clustering approach of Cai and
Heidemann~\cite{cai2011understanding} to network prefixes.  At any rate we
are eager to investigate other data sets, additional protocols and distribution patterns for longer
periods of time.%, as well as effects of BGP churn.
%to analyze prefixes as they are advertised to BGP tables.

% Local Variables:
% mode: latex
% coding: utf-8
% TeX-master: "imc2016"
% End:

%% file: conclusion.tex
\section{Conclusion}
\label{sec:conclusion}

Fast Internet-wide scanning is an emerging topic for investigators who wish
to conduct network research based on up-to-date real world data.  This will
likely lead to a proliferation of scanning activities.  Projects such as
\emph{censys.io} already help to curtail the resulting scan traffic by
making current datasets available to the Internet community for research
purposes.  However, there will always be objectives that call for individual
data collection.  The activities of corporations and individuals must be
factored in as well because tools for fast Internet scanning are widely
available.  It is desirable to research and develop tools that tax the
address space and the patience of scan targets more sparingly than brute
force.  With TASS, we hope to make progress towards the right direction: a
scanning strategy that is more efficient, without loosing significant
accuracy of the results.

Our initial investigations are promising.  By selecting prefixes for
periodic scanning according to density and by adjusting host coverage, it is
feasible to address a wide range of trade-offs.  Particularly, small
compromises with regard to host coverage can reduce scan overhead
substantially, for four protocols that we investigated thus far (FTP, HTTP,
HTTPS, and CWMP).  TASS opens up a variety of options for further research.
When IPv6 becomes popular, brute forcing the address space becomes
infeasible.  By then we ought to have better approaches for network
scanning.  Perhaps TASS can offer a blueprint for tackling that challenge as
well.

% Local Variables:
% mode: latex
% coding: utf-8
% TeX-master: "imc2016"
% End:

%% file: acks.tex
%\newpage
\subsection*{Acknowledgements}
We thank the anonymous reviewers, our shepherd, Roya Ensafi, and Dave Plonka for 
their helpful comments.
We would also like to thank the \emph{censys.io} team for having provided the measurement
infrastructure and for their timely replies to our inquiries.  Furthermore,
we thank Carsten Sch\"auble and Christian Salzmann from the IT service of
the CS department of our university for the compute power they provided for
our analysis effort.  This work was partly funded by grants from the German
Federal Ministry of Education and Research (BMBF Grants No.~16KIS0254,
16KIS0528K).

% Local Variables:
% mode: latex
% coding: utf-8
% TeX-master: "imc2016"
% End: